\documentclass[12pt]{iopart}
\usepackage{graphicx}
\usepackage{color}
\newcommand{\csch}{\mathrm{csch} \,}
\begin{document}
\bibliographystyle{jpcm} 

\noindent
New Journal of Physics {\bf 16} (2014) 023012; doi: 10.1088/1367-2630/16/2/023012

\title{One-dimensional potential for image-potential states on graphene}

\author{P L de Andres,$^1$ P M Echenique,$^{2,4,5}$ 
D Niesner,$^3$ Th Fauster,$^3$ and A Rivacoba$^{2,4,5}$}

\address{$^1$Instituto de Ciencia de Materiales de Madrid (CSIC), Cantoblanco, 28049 Madrid, Spain}
\address{$^2$Donostia International Physics Center (DIPC), P. Manuel de Lardizabal 4, 20018 Donostia, Spain}
\address{$^3$Lehrstuhl f\"ur Festk\"orperphysik, Universit\"at Erlangen-N\"urnberg, 91058 Erlangen, Germany}
\address{$^4$Materialen Fisika Saila, Kimika Fakultatea, UPV/EHU, 1072 P.K., 20080 Donostia, Spain}
\address{$^5$Centro de Fisica de Materiales, Centro Mixto CSIC-UPV/EHU,
Paseo Manuel de Lardizabal 5, 20018 Donostia-San Sebastian, Spain}

\ead{pedro.deandres@csic.es}

\date{\today}

\begin{abstract}
In the framework of dielectric theory the static non-local
self-energy of an electron
near an ultra-thin polarizable layer 
has been calculated and applied
to study binding energies of image-states near free-standing graphene.
The corresponding series of eigenvalues and eigenfunctions have been
obtained by solving numerically the one-dimensional 
Schr{\"o}dinger equation. 
Image-potential-state wave functions accumulate most of their probability outside
the slab. 
We find that 
a Random Phase Approximation (RPA) for the non-local dielectric
function yields a superior description for the potential inside the slab,
but a simple Fermi-Thomas theory can be used to get a reasonable 
quasi-analytical approximation to the full RPA result that can be computed
very economically. 
Binding energies of the image-potential states follow a pattern close to
the Rydberg series for a perfect metal with the addition of
intermediate states due to the added symmetry of the potential. 
The formalism only requires a 
minimal set of free parameters; the
slab width and the electronic density. The theoretical calculations are compared
to experimental results for work function and image-potential states obtained
by two-photon photoemission.
\end{abstract}


\noindent {\it keywords}: graphene, ultra-thin slab,  induced potential,
self-energy, non-local dielectric response,
Random Phase approximation, Fermi-Thomas approximation, 
image-potential states, binding energies, Rydberg series, Whittaker series, work function.

\maketitle

\section{Introduction}\label{intro}
Graphene layers display a number of interesting properties
and potential applications
owing to the linearly-dispersing bands found near the 
$\overline{\mathrm{K}}$ point in the Brillouin zone \cite{novoselov04,castroneto09}.
However, 
in order to get a complete characterization of graphene other
regions in the Brillouin zone need to be considered.
In particular, unoccupied states in the vicinity of the
$\overline{\Gamma}$ point
can play a significant role in the transport of 
currents\cite{C3NR03167E}.
Indeed, it is well known the importance of conduction
band minima valleys located around 
$\overline{\Gamma}$ in ballistic
electron emission processes, where currents are injected
on a substrate under voltage-dependent matching restrictions (e.g.
$k_{\parallel}$-conservation\cite{deandres01})
that are relevant
to design field emission transistors\cite{dubois07}.
On the other hand, the transport of heat also involves
other regions of the Brillouin Zone apart from the 
$\overline{\mathrm{K}}$ point. Recently it has been shown
how the thermal conductivity of few layers of graphene
supported on silicon dioxide depend crucially on flexural modes\cite{seol10}
that are sensitive to the electronic structure around 
$\overline{\Gamma}$\cite{deandres12}.
Therefore,  the dielectric response of very few layers of graphene 
is a key physical element to effectively design
devices based on graphene.
The dielectric response is directly related,
and therefore can be investigated, by trapping
electrons in the region of unoccupied states between the 
Fermi level and the vacuum level. 
These states, bound by their self-induced long-range image potential,
are called image-potential states \cite{echenique78}. 
The experimental \cite{Niesner2012,hofer12},
and theoretical \cite{silkin09}
study of image-potential states
constitutes an ideal probe to better understand the 
properties of graphene layers. 

The image force is a non-local effect asymptotically 
dominated by correlation effects \cite{GMFF}.
In order to study the infinite Rydberg series arising from the
image potential one needs to compute an effective one-dimensional
potential, $\Phi(z)$, representing the real part of the
quasi-static self-energy for an external probe charge. 
This self-induced potential is a continuous function spanning 
from inside the material, where it represents the exchange and
correlation energy, to the vacuum region, where
it should have the correct Coulomb-like asymptotic behavior $\propto-\frac{1}{4z}$.
Such a potential can only be obtained from a non-local spatial formalism,
since a local approach results in a correlation
potential decaying exponentially in the vacuum region,
following the density behavior outside the solid \cite{langYkohn70}.
For a self-consistent first-principles theory such a
non-local functional dependence requires costly numerical
calculations \cite{alvarellos07}.
Therefore, it is useful and natural to search for simpler ways 
to obtain such an effective potential, which is
the basic ingredient needed to understand 
the physics of image-potential states
bound by an ultra-thin polarizable layer like graphene.
The simplest of these alternatives is to introduce a set of fitting parameters to continuously join solutions valid either inside or outside the solid. This point of view has been taken, e.~g.\ by Silkin {\it et al.}\ to study image-potential states in free-standing graphene, joining a function with the correct classical asymptotic behavior outside to a first-principles calculation inside a graphene layer \cite{silkin09}. Such {\it ab-initio} calculations depends on choosing a model for the exchange and correlation potential;  Local Density Approximation (LDA) has been employed by Silkin {\it et al.}\ \cite{silkin09}. Furthermore, it needs setting a very large unit cell to minimize effects between charged periodic images (i.~e.\  $85$ {\AA} vacuum separator was used to reach convergence, implying a large number of plane-waves and a serious computational effort). Finally, such a calculation needs to be supplemented  by a few adjustable parameters, e.~g.\ the choosing of a matching point to join the inside potential to the asymptotic classical potential.

In this paper we analyze an alternative that makes use of a
minimal free parameter set and makes a simple, flexible and accurate
basis for interpreting experimental results.
We use a well-known model for the reflection of
electromagnetic fields at the surface
(infinite barrier specular 
model \cite{ritchie66}),
coupled with a  
non-local static dielectric response
so the desired self-energy can be obtained \cite{deandres87}.
For the electrodynamics model, two free
parameters are introduced, i.~e.\ the 
electronic polarizability of the thin slab, which is
determined by the 
Fermi-Thomas screening wavelength inside the slab
(electron density of the material), and a geometrical dimension
given by the layer thickness.
This approach leads in a natural way to a potential
with proper physical features: it is continuous and finite over the
full spatial domain and it has the right asymptotic behavior towards the vacuum region.
Recently, Ghaznavi et al.\cite{ghaznavi10} have studied the non-linear
screening of a external charge near a doped graphene layer
by solving a Fermi-Thomas model via a non-linear
integral equation.
The non-linear image potential shows changes up to $0.2$ eV
with respect to the classical image potential, and supports the
use of RPA dielectric response for graphene.
In the case of a graphene layer laying on a metallic support
another free parameter may be introduced in this model: the wave-function penetration
in the material. 
Image-potential states are supported in many metallic surfaces for
energies between the vacuum level and the Fermi level due to the existence of
surface band gaps that prevent the penetration of the wave function towards the
bulk and allows the existence of bound states \cite{echenique78}.
Therefore, penetration of wave functions is determined by 
the band structure of particular materials and surface orientations in a way
that cannot be incorporated in our model, except by including a free parameter
that globally determines this penetration.
Here we shall first discuss two limiting cases: 0 or 100\% penetration of the
graphene layer, and then the continuous evolution from one limit to the other.
Since for image-potential states we are interested in regions in reciprocal space with
$\vec k$ near $\overline{\Gamma}$ and
energies between the vacuum level
and the Fermi energy,
it is well justified to model graphene as a polarizable 
electron gas with a quadratic energy dispersion, as seen from the 
relevant bands for 
graphite and graphene\cite{fauster83,posternak83, posternak84,silkin09}.
A connection between the interlayer states and the image potential states
on graphite, graphene and other carbon-based materials has
been established by different groups, e.g. Silkin et al.\cite{silkin09}, 
Feng et al.\cite{feng11}, and Hu et al.\cite{hu10}.
These authors have pointed out that such states exist universally on
two-dimensional materials. 

The static dielectric response, $\epsilon(\vec k)$,
has been modeled by a Random Phase
Approximation (RPA), and by its small $k$ expansion,
the Fermi-Thomas Approximation (FT) \cite{pines}.
While the RPA yields a more accurate description of excitation
in the material, 
introducing the FT allows us to write the potentials as analytic
expressions or quasi-analytical ones which merely depend 
on a final numerical step involving
the simple integration of a function decaying quickly
for large values of the argument.
A single parameter,
the screening constant related to the density of
states at the Fermi level, $k_\mathrm{FT} \propto \frac{\partial n_0}{\partial \mu}$,
fixes both the scales for energies and lengths
facilitating the rescaling of results for different materials and sizes
(atomic units are used throughout the paper, except where
explicitly stated otherwise). 
Taking graphite as a model 
(2 g/cm$^{3}$, 2s$^2$ 2p$^2$), a typical value for graphene 
is $k_\mathrm{FT} \approx 1$ ($r_s \approx 2.5$),
although its precise value should depend on factors like doping,
external potentials, etc; this is accommodated in our results through
the simple aforementioned scaling with $k_\mathrm{FT}$.
The other parameter used to characterize a thin slab is its
width, $2d$.
For a single atom thick layer of graphene a reasonable value for $d$,
should be related to the spatial extension of $\pi$ carbon orbitals,
$d \approx 1$ a.u.
The value of this parameter turns out not to be critical for
this work because wave functions spread over regions
much larger than $d$.

The theoretical results are compared to experimental results for graphene on
various substrates. The presence of the substrate leads to a charge transfer 
from or to the graphene which can be described also as doping. The resulting
work function change due to graphene has been modeled by Giovanetti {\it et al.}\
\cite{Giovannetti2008,Khomyakov2009}. We find a good agreement with the experimental data.
The doping changes the available screening charge and leads in turn to
a change in the energy of the image-potential states. The experimental data
from two-photon photoemission agree with the calculated dependence.

\section{Theoretical model}\label{theo}
\begin{figure}
\includegraphics[clip,width=0.9\columnwidth]{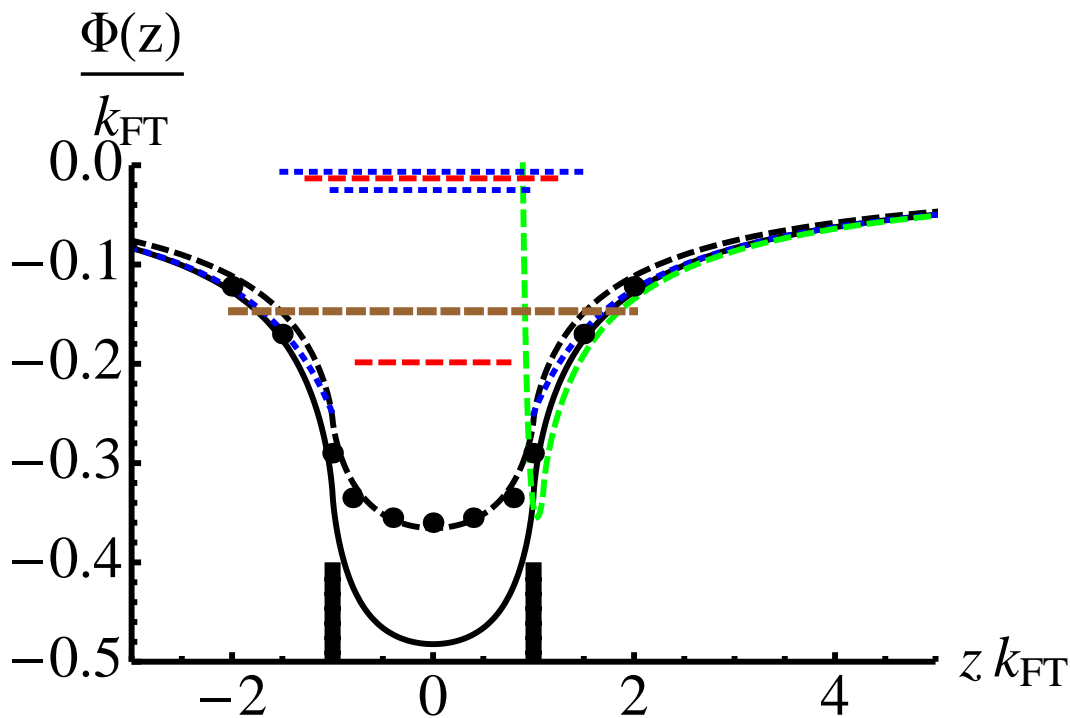}
\caption{Self-energy, $\Phi(z)$,  
of a unit test charge at different positions
outside and inside a thin slab  
($k_\mathrm{FT}=1$ and $d=1$).
The two thick vertical lines (black) mark the slab region.
Black continuous line 
corresponds to FT (Eqs. (\ref{eq:FTvI}) and (\ref{eq:FTvII})),
while black dots have been obtained using the RPA.
Black dashed line: FT approximant to RPA
($k_{FT}^{*}=0.78$).
Blue dotted line: asymptotic law with image
plane at $z_0=-\frac{1}{k_\mathrm{FT}}$ ($\mid z \mid > d$).
Dashed line (green): 
$\Phi(z)+e^{-20(z-b)}$ (repulsive term) for $b=d$.
Horizontal dashed and dotted lines 
show the first four eigenvalues (red and blue for
even and odd, respectively);
the horizontal thick line (brown) gives an approximate
value for the work function in graphene ($4$ eV).
}
\label{fgr:LAM28}
\end{figure}

\subsection{Self-induced potential by an ultra-thin slab}\label{potential}
For an external probe charge near a slab ($Q=1$) we seek the potential
acting on $Q$ by the polarization charges induced in the medium
by $Q$ itself.
This is obtained by computing the total potential,
and subtracting the charge's own naked potential.
To ensure the proper boundary conditions, and according to 
the specular reflection model at the surface, auxiliary
pseudo-media are introduced for the polarizable slab and the
vacuum that reduce the calculation to matching solutions obtained
in different regions of space for homogeneous media 
everywhere \cite{GMFF}.
Details of the calculation for the thin slab are given in \ref{Slab}.
The resulting potential depends on a number of integrals that
include the dielectric response of the system: for the RPA
these are computed numerically. On the other hand, 
within the FT approximation
an expression that only depends on a single numerical 
integration can be obtained,

\begin{equation}
\Phi_\mathrm{FT}(z > d)=-
\frac{k_\mathrm{FT}^2}{2}\int_{0}^{\infty }
 \, d \kappa \,
\frac{e^{-2 \kappa z} }{ \left(
\chi+
\kappa \coth{
\left[
\chi d \right]}\right)
\left(\chi+\kappa
\tanh\left[\chi d \right]\right)}
\label{eq:FTvI}
\end{equation}
\noindent

\begin{equation}
\label{eq:FTvII}
\Phi_\mathrm{FT}(0<z \le d)=
\end{equation}
$$
\int_{0}^{\infty } \, d \kappa \bigg \lbrace 
\frac{\chi+e^{4 \chi d} 
\left(\kappa e^{2 \chi z }+\kappa -\chi \right)}
{2 \left(
e^{4 \chi d}
-1\right) \chi}
+
\frac{\kappa}{2 \chi}
\bigg \lbrack
-
\frac{2 \kappa 
 \left(
(\chi+\kappa)
e^{2 (2 d+z)\chi }
+
(\chi-\kappa)
\right)}
{(2 \kappa^2+k_\mathrm{FT}^2)
\left(e^{4 d \chi}-1\right)  +
2 \kappa \chi
\left(e^{4 d \chi}+1\right)}
+
$$

$$
+
\frac{e^{-2 \chi z}+1}{e^{4 \chi d}-1}
-
\frac{e^{-2 \chi (d+z) } 
\left(1+e^{2 \chi z}\right) \kappa 
\left(\kappa+\chi \left(e^{2 (d+z) \chi
}+\cosh \left[2 \chi d \right]\right) 
\csch \left[2 \chi d \right]\right)}
{2 \kappa \chi
\cosh\left[2 \chi d \right]+
\left(2 \kappa^2+k_\mathrm{FT}^2\right) \sinh \left[2 \chi d \right]}
\bigg \rbrack
\bigg \rbrace
$$

\noindent
where $\chi=\sqrt{\kappa^2+k_\mathrm{FT}^2}$.
This is an useful expression that can be
computed very efficiently. We shall see that 
for the purpose
of computing the energy levels of image-potential states
it makes an excellent approximation
to the more costly RPA calculation. 

In figure~\ref{fgr:LAM28} we show the potential
for a slab occupying the region $-d \le z \le d$;
both in the FT approximation (black continuous line),
and in the RPA one (black dots).
In the region determining the Rydberg series
($|z|\ge d$), both approaches yield similar values
and agree with the correct asymptotic power-law.
Near the center of the slab, 
FT overestimates the interaction over RPA
by about 30-40\%, 
\begin{equation}
\frac{\Phi_{RPA}}{\Phi_\mathrm{FT}} \bigg|_{z=0} \approx 0.82-\frac{r_s}{12.5}
\quad ; \quad 2 \le r_s \le 6
\end{equation}\label{eq:CH}
\noindent
a difference that is reflected mainly in the lowest state
(node-less)
that has a significant weight in the central part of the slab
where the difference between RPA and FT is larger. 
This state is located outside the window between the
vacuum level and the Fermi energy and
is not part of the image-potential states series.
For $d\gg\frac{1}{k_\mathrm{FT}}$ the $\kappa$ integral 
in Equations~(\ref{eq:FTvI}) and ~(\ref{eq:FTvII}) can be evaluated
analytically to obtain the following useful particular values
(see \ref{Asymptotics}):
$\Phi_\mathrm{FT}(z=d)=-\frac{k_\mathrm{FT}}{3}$,
and $\Phi_\mathrm{FT}(z=0)=-\frac{k_\mathrm{FT}}{2}$. 
We remark that the latter 
value corresponds to the Coulomb hole \cite{hedin99}.
Furthermore, 
for $z > \frac{1}{k_\mathrm{FT}}$ (vacuum region)
the potential is well approximated by
a classical law,
$\frac{1}{4(z-z_0)}$,
corrected by 
an image plane, $z_0$.
The value of
$z_0$ can be obtained by expanding the integrals for
$\kappa \approx 0$, which fixes the position of the
image plane in this model: $z_0=-\frac{1}{k_\mathrm{FT}}$
(dashed blue line in figure~\ref{fgr:LAM28}, see \ref{Asymptotics}).
Finally, Equation~(3) suggests a simple procedure to find
a good approximant to the RPA potential. Given a 
physically representative value for
$k_{FT}(r_s)$, one can find another value, $k_{FT}^{*}$,
such that $\Phi_{RPA}(k_{FT},d=0) \approx \Phi_{RPA}(k_{FT}^{*},d=0)$.
For the purpose of obtaining energy eigenvalues of
$\Phi_{RPA}(k_{FT},z)$,
$\Phi_{FT}(k_{FT}^{*},z)$ is a good 
approximation to the full RPA potential
that can be computed quick and easy.

\subsection{Eigenvalues}\label{eigenvalues}

We solve numerically the Schr\"odinger equation  \cite{crandall82,shooting} to 
compute the eigenvalues and eigenfunctions 
corresponding to the RPA and FT model
potential described above.
Quite generally, bound states can be 
unequivocally labeled by the number of nodes $n$,
with energies increasing as the number of nodes increases.
Furthermore, as long as the potential is symmetric, eigenfunctions have
either even or odd parity, for even or odd number of nodes, $n$.

\begin{table}
\begin{tabular}{r|r|rrrrrrr
}
\hline \hline \hline \hline
label      & 0 &  1 & 2 & 3 & 4 & 5  & 6  & 7 \\ \hline \hline
RPA    &-5.38 &{\bf -.68}& -.36&{\bf -.18}&-.14&{\bf -.090}& &
\\
FT     &-7.14&{\bf -.89}& -.44&{\bf -.22}&-.15&{\bf -.096}&-.074&{\bf -.054}
\\ \hline
LDA ($z_0=5$)\cite{silkin09}  & &{\bf -1.29}   &-.57& {\bf -.24}&-.17& {\bf -.11}&-.06
\\ 
$R_{b=-20}$&  -6.03   & {\bf -.79}  & -.41 & {\bf -.19}  &  &   &&
\\ 
\hline \hline \hline \hline \hline
label    &   & 1 &    & 2  &    & 3  &    & 4
\\ \hline
$-\frac{1}{4z}$\cite{echenique78} &    & {\bf -.85}&   &{\bf -.21}&    &{\bf -.094}&    &{\bf -.053} 
\\ 
Gr/Ir\cite{Niesner2012} &    &  {\bf  -.83}&  & {\bf -.19} &  & {\bf -.09} &       
\\ \hline
\hline \hline \hline \hline
$R_{b=1}$  &               &{\bf -.57}   &                  & {\bf -.17}  & & {\bf -.082}  & & {\bf -.048}    
\\ 
$-\frac{1}{4(z+z_0)}$ ($\delta=0.2)$ &    & {\bf -.59}&   &{\bf -.18}&    &{\bf -.083}&     & {\bf -.048}
\\ \hline \hline \hline \hline
\end{tabular}
\caption{\label{tbl:BE}
Binding energies, $E_{n}$ (eV), measured with respect to the vacuum level
and labeled according to 
the number of nodes in wave functions.
Top three rows:
Results for RPA and Fermi-Thomas (Eq. \ref{eq:FTvII})
are given for a free-standing graphene ultra-thin layer 
and compared with similar results obtained by 
from an {\it ab-initio} LDA calculation matched to an
asymptotic expression for the image potential by Silkin et al.\cite{silkin09}
(states are labelled accordingly to
their ordering in the image-potential series of unoccupied states
lying above the Fermi energy and below the vacuum level).
The effect of a repulsive barrier located at $z \approx -10$ {\AA} is
presented under $R_{b=-20}$ (notice that in this case symmetry
is broken and the parity is no longer a good quantum number).
For comparison, experimental values measured on Gr/Ir are quoted\cite{Niesner2012},
and compared with the eigenvalues for the classical Whittaker's problem\cite{echenique78}.
Finally, we show the effect of approaching the repulsive barrier to $z \approx 0.5$ {\AA},
$R_{b=1}$,
that can also be computed by introducing an appropriate quantum defect ($\delta=0.2$).
To guide the eye, we highlight in bold face numbers that can be compared across different
calculations or experiments and have been given an accompanying interpretation
in the text.
}
\end{table}

We show in 
figure~\ref{fgr:LAM28} 
the first five 
eigenvalues for $\Phi(z)$; dashed (red) and dotted (blue) horizontal lines for
even and odd parities. 
It is worth noticing the structure of this series:
there is an isolated eigenvalue (the lowest one), while
the remaining states cluster near the vacuum level.
For standard densities (e.~g.\ $1 < r_s < 10$) this first level
appears below $\sim -4.5$ eV; an estimate for the work function
in graphene (thick line).
Therefore, this $n=0$ eigenstate does not fit in 
the standard definition for 
a Rydberg state, necessarily located between the vacuum
and Fermi levels to be observable in a standard
experiment. 

\begin{figure}
\includegraphics[clip,width=0.9\columnwidth]{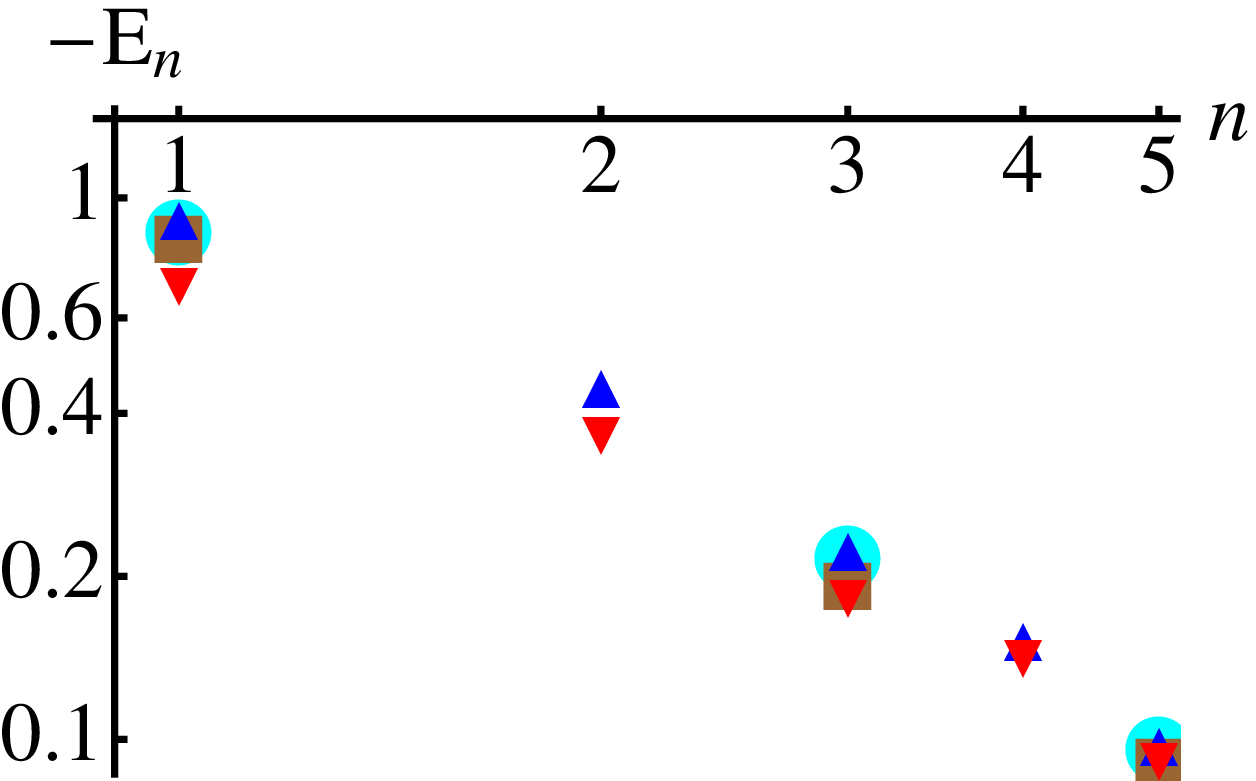}
\caption{Bound states energies, $-E_{n}$ (eV), for the
hydrogenic series supported by the potential of the 
ultra-thin slab in figure~\ref{fgr:LAM28} 
(FT: triangles up, blue; RPA: triangles down, red).
These are compared with 
experimental values for 
Gr/Ir (squares, brown) \cite{Niesner2012},
and Whittaker series (circles, cyan).
Energies have been labeled on the abscissa by the
quantum numbers $n$
corresponding to wave functions from the symmetrical potential
(Whittaker quantum numbers have been accordingly transformed to $n=2 m - 1$,
$m=1, 2, 3, \dots$).
}
\label{fgr:autoE}
\end{figure}

In table~\ref{tbl:BE} we compare eigenenergies calculated for the RPA and FT models
with similar theoretical values obtained from a formalism based in Density Functional Theory\cite{silkin09}, and with experimental values reported for Gr/Ir \cite{Niesner2012}. We have also included, as a reference, the limiting case of the Rydberg series for a perfect metal $E_{m+1}=-\frac{1}{32 (m+1)^2} ; \, m= 0, 1, 2, ...$, where $m$ refers to the number of nodes for each state (note that these wave functions only extend to $z>0$ half-space, and that the zero at the origin is not counted as a node since it derives from the boundary conditions).
Figure~\ref{fgr:autoE} shows in a log-log scale the 
scaling law for the first few members of the series.
We remark that for the purpose of computing binding energies for
image-potential states the FT model 
with a suitable value for $k_{FT}^{*}$
yields values of the same quality as the
RPA at a much lower computational cost.
For the potential created by free standing graphene Silkin et al.\cite{silkin09} have computed binding energies for image states by matching an {\it ab-initio} Density Functional Theory (DFT) model for the total energy to an asymptotic expression for the image  potential beyond some distance, $z_0$. A pseudopotential with valence electrons 2s2 2p2 and a pseudized core for 1s2 electrons has been used. In order to compare results from these different theoretical approaches we align the states attending to their energies in the same order as they appear in the image series. This labeling of states is straightforward and unambiguous, and most importantly it facilitates comparison between different theories and with experimental values. On the other hand, theoretical eigenvalues can be classified by looking at the properties of their eigenfunctions, notably the number of nodes and for a symmetric potential the parity. Eigenvalues corresponding to a symmetric potential like the one plotted in Fig. 1 are easily classified in the same order as they appear, $n$,  by the number of nodes (n), and its parity ($+$ for even $n$, and $-$ for odd $n$). However, parity might become not a good quantum number in the presence of defects or a perturbing potential modifying the symmetry of the potential (e.g., a supporting substrate for the graphene layer), while the ordering dictated by the number of nodes is related to the orthonormality of wavefunctions and makes a robust labeling scheme. Within this ordering, the first image-potential state is simply the first unoccupied one to lie above the Fermi energy and below the vacuum level determined by the work function, $W$. In the RPA/FT model we are introducing in this paper the first image-potential state has been labeled $n=1$ in table~\ref{tbl:BE}, its wavefunction has one node and it is antisymmetric, $E(1)_{RPA}=-.68$ eV.  In the DFT model, the first image-potential state we find is $E(1)_{DFT}=-1.29$ eV, and its wavefunction has two nodes and it is symmetric (labeled in ref. \cite{silkin09} as $1^{+}$).  The number of nodes on this wavefunction has been correlated in ref\cite{silkin09} with the occupied $\sigma$ and $\pi$ states below and the required orthonormality condition between them. The discrepancy between the number of nodes and parity on wave functions from these two theoretical approaches is not a fundamental one. Presumably a pseudopotential including a different core, like a simpler one constructed with only 2p2 electrons, or a more complete one including the 1s2 electrons, would alter the number of nodes in the first unoccupied state because the orthonormality condition to the low lying states would be different. Similarly, in our RPA model one could imagine an scenario where the value of $k_{FT}$ is so small as to make the $n=0$ state the first unoccupied state, or inversely, so large as to get the $n=1$ state below the Fermi level to make the $n=2$ state the first member of the series of unoccupied states. Therefore, we conclude that the more convenient way to compare different image series from different theories, or with experiments, is the ordering attending their energy, because it does not depend on the fine details of the model being used. This criterion brings good agreement between DFT and RPA/FT, and with experiments, which is remarkable taking into account how different are both theories, and something to be highlighted. Perhaps the exception is the first value predicted by DFT, $-1.29$ eV, that is a bit too low. We do not take this small discrepancy as serious since both theories involve a number of approximations and parameters (the pseudopotential and the exchange-correlation potential for DFT; $k_{FT}$ and $d$ for RPA), and it is well known the difficulties in DFT to get accurate values for empty states (e.g., band gaps) without including a more sophisticated description for electron correlation, which is the physical effect responsible for the asymptotic behavior, and one of the good assets in the RPA/FT formalism.

\subsection{Eigenfunctions}\label{eigenfun}

\begin{figure}
\includegraphics[clip,width=0.45\columnwidth]{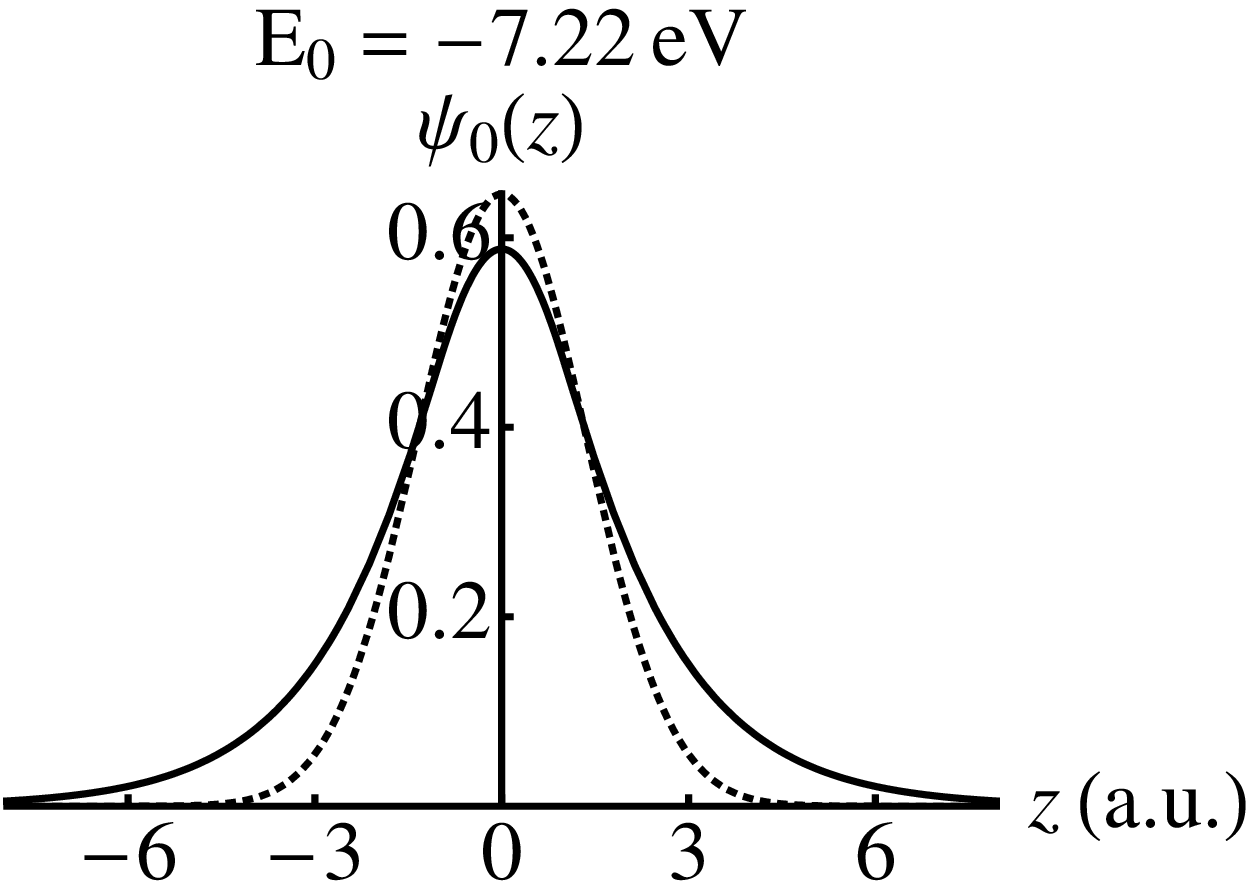}
\includegraphics[clip,width=0.45\columnwidth]{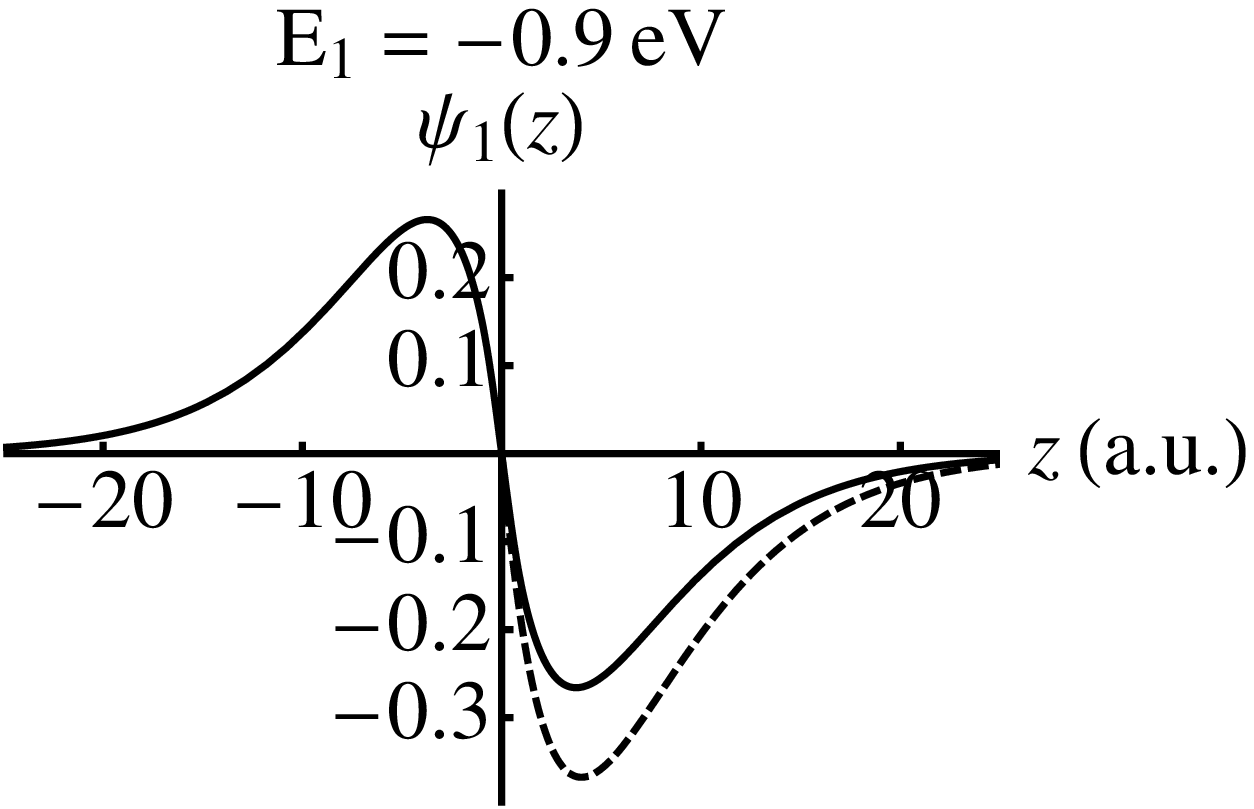} \\
\includegraphics[clip,width=0.45\columnwidth]{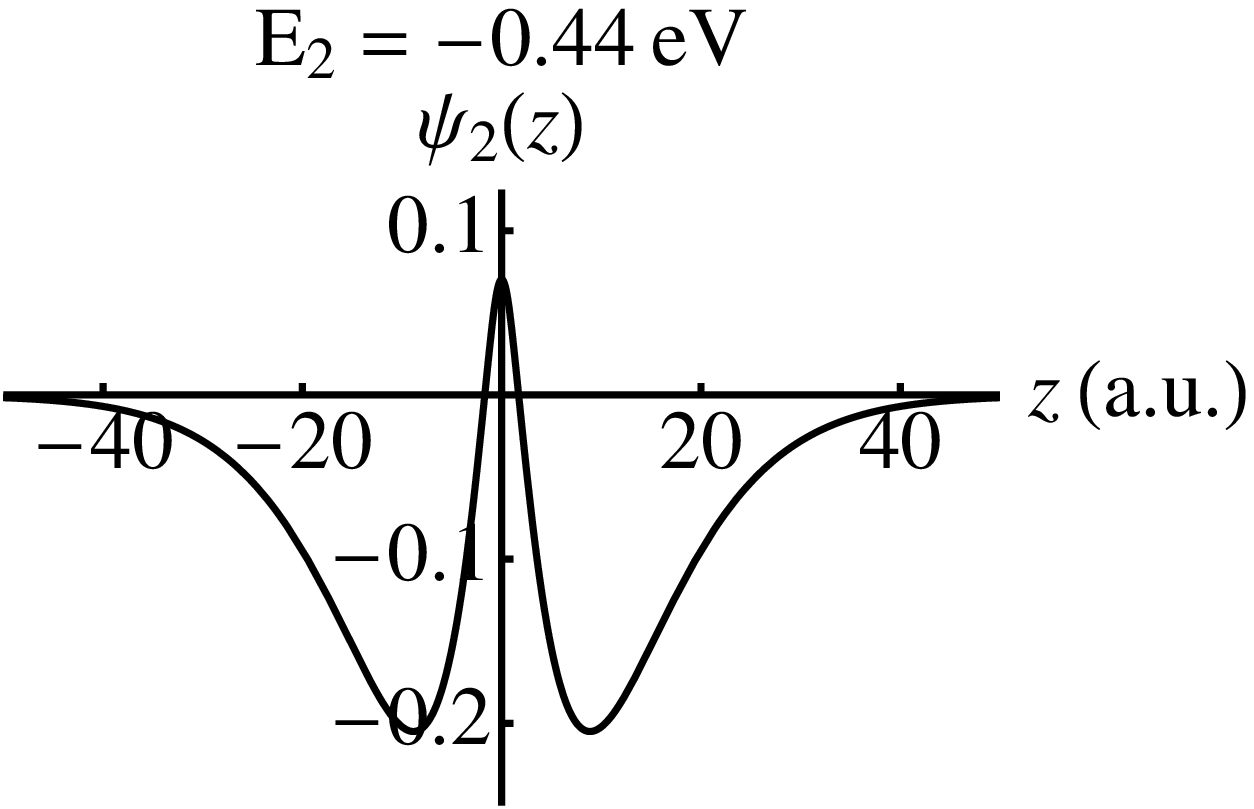}
\includegraphics[clip,width=0.45\columnwidth]{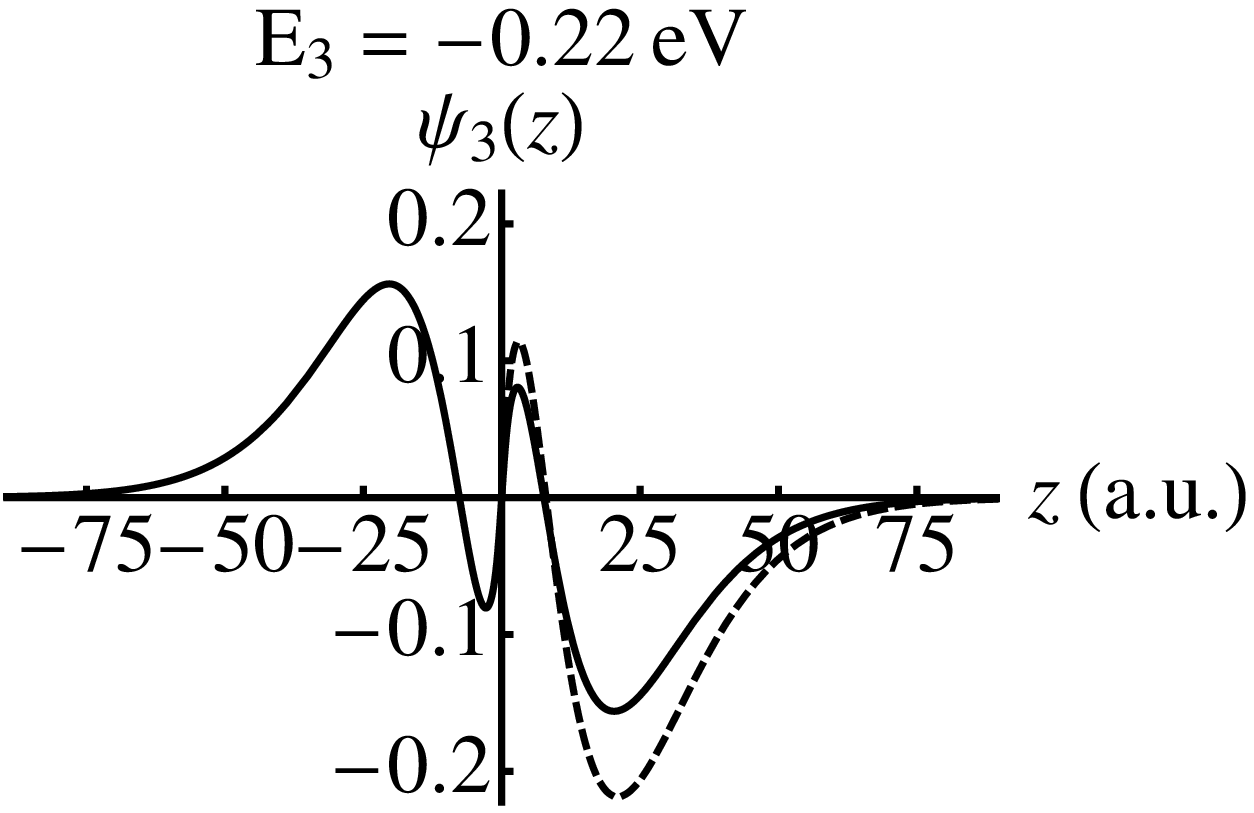}
\caption{
First four eigenfunctions for the potential displayed 
as a black continuous line in
figure~\ref{fgr:LAM28} (FT). 
Eigenvalues are in eV and referred to the vacuum level.
For comparison, Whittaker wave functions ($n=1, 3$, dashed),
and the fitted harmonic oscillator wave function
($n=0$, dotted) are shown.
}
\label{fgr:autoF}
\end{figure}

The similarity of the values found for the antisymmetric (odd, $n^{-}$)
members of the series of states for $\Phi(z)$ and the classical Rydberg series 
is striking, and merits some attention (third row in table~\ref{tbl:BE}).
Such a similarity can be understood by looking at the
corresponding wave functions (figure~\ref{fgr:autoF}).
While wave functions for Rydberg states
are spatially located mostly in the vacuum region, 
the symmetric members of the series have
$\psi(z=0) \ne 0$ at the origin of the slab
(figure~\ref{fgr:autoF}, continuous
line in upper-left panel).
This is most conspicuous for the ground state wave function $\psi_{0}$
that  is more alike
to the ground state of a harmonic oscillator fitted
to the bottom of the potential well 
($\Phi(z)\approx -0.5 + 0.13 z^2$ a.u.,  $E'_0=-6.2$ eV),
than to
states in the one-dimensional hydrogen-like series for a semi-infinite metal, that
are assumed to go to zero at the image plane.

Boundary conditions force all Whittaker wave functions 
to go to zero at the origin, 
a condition that in the case of a symmetric well can
only be fulfilled by odd wave functions.
Moreover, if $n^{-}$ and $m$ give the number of nodes for odd wave functions 
for the symmetric potential, and the Rydberg one respectively, we
can make a one-to-one correspondence, $\frac{n^{-}-1}{2}=m$, 
that simply tells us that both sets of wave functions have the same number
of nodes if $n^{-}$ is divided by two (only half-space) and the node
at the origin is discounted.
To compare with wavefunctions corresponding to Whittaker's classical problem
defined only over half the space one might use Silkin's et al. labelling 
for image states: a quantum number related to nodes
of wavefunctions for the symmetrical problem 
only on half of the space supplied with its parity to
take into account that antisymmetric wavefunctions
add an extra node to the count. 
From a physical point of view, we can envisage two relevant limits:
A free standing slab producing a symmetric potential with states labeled
by the number of nodes and their parity, 
and a slab on a substrate where a particular surface gap may prevent penetration
of wave functions inside the material leaving only half the space accessible for
image-potential states.
In this later scenario, parity would not be an useful quantum number.
It is reasonable to describe the case of a graphene layer grown
and supported on a particular substrate as having a place somewhere in
between these two limits depending on details like the interaction
between the graphene layer and the support, the electronic surface structure
of the combined system, work function, etc. 
In what follows, we shall assume that all these details
can be taken into account in the simplest terms by a single free parameter
giving the amount of penetration of wave functions.
This electrodynamics formalism is not intended to describe these details, 
but it can bring useful quantitative information on the observed levels,
and as a consequence to further conceptual understanding. 

\begin{table}
\begin{tabular}{c|rrrrrrrrrr}
\hline
Whittaker           & 0    &      & 1    &      & 2    &      & 3    &      & 4 \\ 
$\overline{z}$      & 3    &      & 12   &      & 29   &      &   51 &      & 79 \\ \hline
RPA (FT) & 1    & 2    & 3    & 4    & 5    & 6    & 7    & 8    & 9  \\ 
$\overline{z}$      & 3    & 6    & 12   & 18   & 28   & 34   & 48   & 58   & 78 \\ \hline
\end{tabular}
\caption{\label{tbl:xM}
Expectation value 
$\overline{z}$ ({\AA})
for Whittaker and RPA wave functions
(FT results are very similar to RPA ones).
}
\end{table}

This formalism predicts, for a free-standing graphene layer,
the appearance of new states associated with the even parity
(e.~g., eigenvalues between $-.44$ and $-.36$ eV for $n=2$, 
and between $-.15$ and $-.14$ eV for $n=4$
in Figure~\ref{fgr:autoE}).
These new states have been obtained numerically 
and fit well into the classical
scaling law proportional to $n^{-2}$ for $n$ up to 7.
Obviously, the symmetric potential can be perturbed by external ones
(e.~g. the cases of graphene on a support), and these states
would be affected accordingly.

\subsection{Modeling the substrate}\label{substrate}

So  far we have discussed a model that effectively represents a
free-standing graphene layer. 
For those cases where the graphene layer has been deposited on
a metallic support the substrate is expected to
manifest itself in two main physical ways:
(i) The wave-functions may be constrained to be outside some spatial region
where the substrate enforces an electronic gap, and
(ii) as a consequence of the interaction between the layer and
the support, some charge may be transferred to/from the slab,
modifying the density of states at the Fermi level, i.~e., the
value of $k_\mathrm{FT}$.

To assess how sensitive the eigenvalues are to the penetration of
wave functions into the material
we have added to $\Phi(z)$ a repulsive term modeled 
as an exponential wall, $R_{b}(z)=e^{-a (z-b)}$.
Since we are only interested in creating a decaying state
inside the support, we fix the parameter $a$ to a large
value, $a= 20$ a.u., akin to the infinite hard-wall limit;
its effective role is to expel states from the $z<b$ region,
ensuring the exponential decay of wave functions inside
an electronic band gap.
The resulting potential for the repulsive barrier located near the
slab surface ($b=1$ a.u., green dashed line in figure~\ref{fgr:LAM28}) is similar
to the classical series with an image plane, 
$\frac{1}{4(z+z_0)}$, and can be
easily solved by introducing a quantum defect in Rydberg formula
(compare energies for the same number of nodes in the 
eigenfunctions for $R_{b=1}$ and Whittaker series with 
quantum defect $\delta=.2$
in table~\ref{tbl:BE}).
The barrier, on the other hand, can be introduced below the
surface,
mimicking the effect of an electronic band gap due to
a supporting substrate.  
The evolution of the first few eigenvalues with the position of
the barrier have been shown in the two limits in table~\ref{tbl:BE}.
As long as the barrier is located far away
from the ultra-thin slab (e.~g.\ $b \le 20$ a.u.) we get values
reminiscent from the original members for the unperturbed
symmetrical potential. 
On the other limit, a barrier located just on the surface very much
reminds of the classical solution.
For large $m$ the eigenvalues are determined by the potential
in the vacuum region and the existence of such a barrier
distorts less and less the states as they approach the vacuum level. 
Table~\ref{tbl:xM} gives the expectation mean values in {\AA},
$\overline{z}=\int_{0}^{\infty} \psi (z) \, z \, \psi(z) \, dz$,
for Whittaker wave functions compared with the ones 
obtained for the RPA or FT potentials (e.~g.\ 
(\ref{VvacFTscib})).
These values compare well with each other, 
which reflect the manifest similarity between wave functions
commented on figure~\ref{fgr:autoF},
and show how the important region for the potential moves
quickly away from the surface as $m$ grows.
The fact that $\overline{z} \gg d$ for image-potential states 
with $n > 1$ implies that wave functions are quite insensitive to
the potential inside or near the layer and
they are mostly influenced by the asymptotic region where FT and
RPA are equivalent.
This suggests that
higher $k$-corrections to the dielectric function arising from
the random phase approximation are not very important, at least for $n > 1$ states.
Taking away the first level, largely affected by the details near the bottom
of the potential,
the rest of the series is only
modified by a percentage comparable
to differences found
in table~\ref{tbl:BE} between similar entries. 

\begin{figure}
\includegraphics[clip,width=0.75\columnwidth]{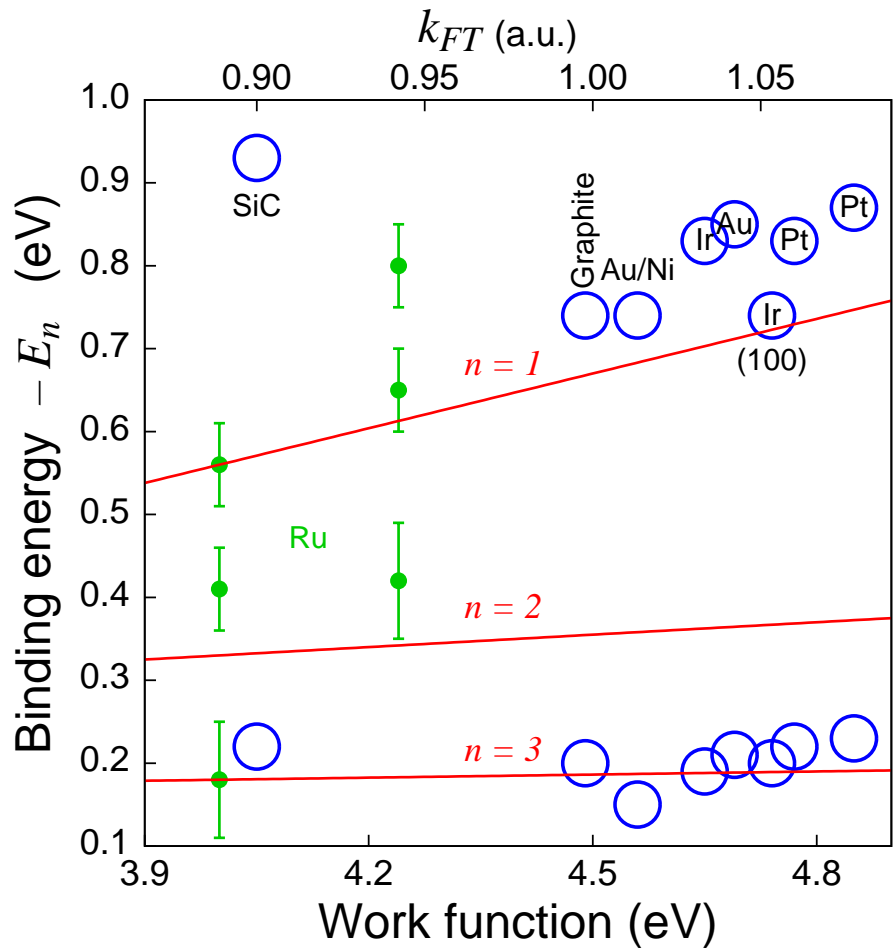}
\caption{
Solid lines (red): Energy of $n=1, 2,$ and 3 image-potential states
as a function of work function, $W=\frac{k_\mathrm{FT}}{6}$ (eV).
Large open circles (blue) show experimental results\cite{Niesner2012,Nobis2013,Gugel2013}. 
For Ru the data\cite{hofer12} are plotted by filled circles (green) for
two different work functions: $4.00$ and $4.24$ eV.
Data for graphite have been taken from\cite{takahashi12}.}
\label{fgr:EnWth}
\end{figure}

The effect of doping may be explored by looking at the 
$k_\mathrm{FT}$ dependence of the energies of the image series.
The important parameter of the dielectric model is the charge density of the graphene layer which can be related to the doping level and the work function. This is summarized in the current formalism via a single parameter, the Fermi-Thomas wave-vector 
$k_\mathrm{FT}$. While $k_\mathrm{FT}$ cannot be easily extracted in our formalism from doping levels of graphene, we can compare to experiments by exploiting the linear dependence predicted by this theory between the screening wave-vector and the work function of the thin slab. 
To compare with available experimental results it suffices to establish
a connection between $k_\mathrm{FT}$ and a relevant energy scale,
e.~g. the distance between the vacuum level to the Fermi level,
i.~e. the work function, $W$. To this end,
we note that within our approach the work function for a semi-infinite 
surface, or a slab with $d>1$ {\AA}, is $-\frac{k_\mathrm{FT}}{2}$
in the Fermi-Thomas model, and $\approx -\frac{k_\mathrm{FT}}{3}$ in
the RPA one. 
We can see that RPA, so far our best approach,
overestimates $W$ when compared with experimental values
by about a factor $\approx 2$.
The linear dependence between $W$ and $k_\mathrm{FT}$, on the other hand,
is solidly anchored in the theory, having its origin in the net attractive
interaction between the external electrons and the polarization charges
created inside the polarizable material.  
Therefore, $W$ should be proportional to $ \alpha k_\mathrm{FT}$, albeit the
proportionality constant should be corrected to compare with
the experimental value. We take $\alpha=\frac{1}{6}$, 
that corresponds to the empirical work function
for graphene, 
$W_0=4.5$ eV. 
Therefore, the empirical dependence $W=\frac{k_\mathrm{FT}}{6}$ 
allows us to translate $k_\mathrm{FT}$ into an experimental energy scale,
and to set up an energy origin at the value $W_0$.
Figure~\ref{fgr:EnWth} shows the results for the first three
members of the series by straight lines. 
The experimental data will be discussed in section \ref{exp}.

\begin{figure}
\includegraphics[clip,width=0.75\columnwidth]{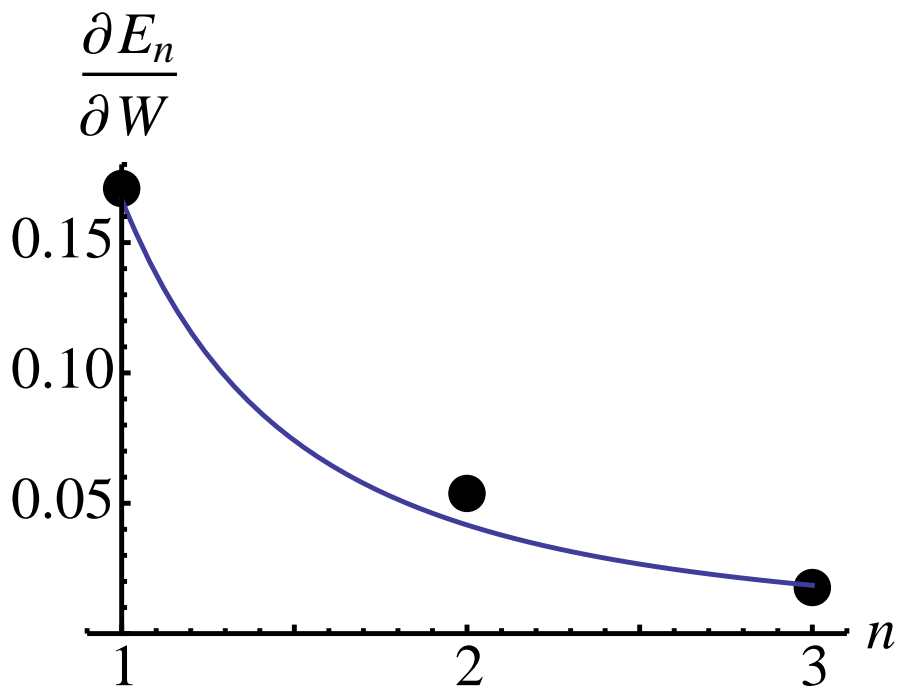}
\caption{
Scaling of the slope of $\partial E_n(W)/\partial W$
computed for the first three image-potential states.
}
\label{fgr:DEnRhoth}
\end{figure}

An interesting feature of this result is the
fact that the higher member of the series are less and less affected
by changes in $k_\mathrm{FT}$ ($W$ and/or $\rho(E_F)$).
This is clearly seen in Figure~\ref{fgr:EnWth} where
the slope of $\partial E_n(W)/\partial W$ decreases for the higher $n$
values. 
In Figure~\ref{fgr:DEnRhoth} we plot this result, and show that it
is well fitted by the empirical function $\frac{1}{6 n^{2}}$.
Since $k_\mathrm{FT}$ is linearly proportional 
to the density of states at the Fermi level, $\rho(E_F)$, this
gives us the empirical scaling law expected for the different
terms of the image-potential-state series with doping.

\section{Experimental results}\label{exp}

The theoretical results of the preceding section can be tested experimentally by
two-photon photoemission (2PPE). This technique is able to measure image-potential states with high accuracy
and results for various graphene-covered surfaces have been reported 
\cite{Niesner2012,hofer12,Nobis2013,Gugel2013}. 
Photoelectron spectroscopy can also provide precise values for the work function of the surfaces
which is correlated to the doping level and charge density of the graphene layer.
We first discuss the work function before turning to the image-potential states.

\subsection{Work function}\label{workfun}
\begin{figure}
\includegraphics[clip,width=0.95\columnwidth]{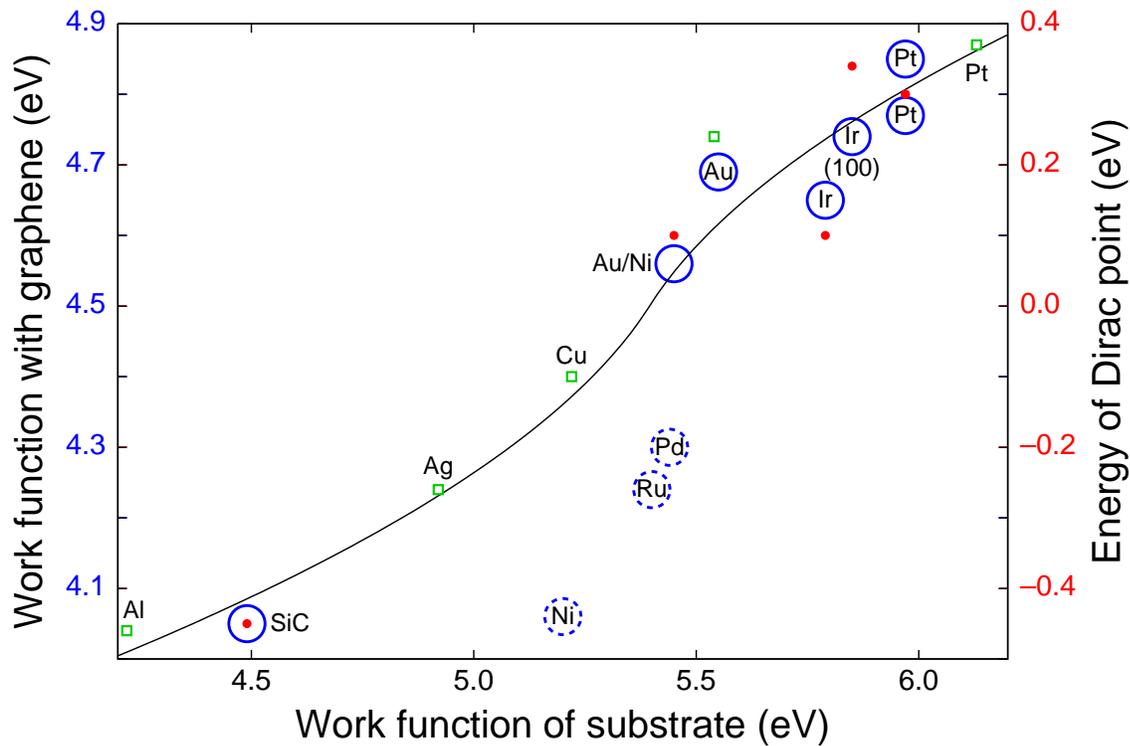}
\caption{Work function of graphene-covered surfaces versus the work function of the clean substrates (blue open circles). The solid curve shows the results of a capacitor model which match the calculated results plotted by green open squares \cite{Giovannetti2008,Khomyakov2009}. 
The energies of the Dirac point (red filled circles) follow the same curve 
with an energy scale shifted 4.5 eV relative to the vacuum level.
}
\label{fgr:workfun}
\end{figure}

The amount of charge transfer from the substrate to the graphene layer is determined by the 
work function difference between substrate and graphene. This situation has been modeled
with a simple capacitor model by Giovanetti {\it et al.}\
and validated by comparison to results from calculations for various surfaces 
\cite{Giovannetti2008,Khomyakov2009}.
The curve plotted in figure~\ref{fgr:workfun} shows the calculated work function
of the graphene-covered surfaces versus the work function of the clean substrates for the capacitor model.  
The calculated values plotted by green open squares fit the curve quite well \cite{Giovannetti2008,Khomyakov2009}.
The available experimental data are shown by blue open circles 
\cite{Niesner2012,hofer12,Nobis2013,Gugel2013,Oshima1997,Murata2010a}. 
The size of the circles represents approximately
typical experimental error bars. The curve was fitted to the experimental results for the noble-metal surfaces
\cite{Nobis2013} and monolayer graphene on SiC \cite{Gugel2013}. 
The fit was done with the work function of graphene of 4.50~eV compared to 4.48~eV in the original work \cite{Giovannetti2008,Khomyakov2009}. 
The chemical shift was reduced from 0.90~eV to 0.89~eV\@. Overall the fit of this work describes also the
calculated values (green open squares in figure~\ref{fgr:workfun}) very well.
The surfaces included in the fit all have a
graphene-metal distance around 3.4~{\AA} \cite{Varchon2007,Martoccia2008,Sutter2009,Busse2011,Slawinska2012}. 
The surfaces of Ni, Ru, and Pd were not included in the fit and shown by blue dashed circles. 
They have shorter graphene-metal distances \cite{Giovannetti2008,Khomyakov2009,Moritz2010} 
and would require a larger chemical shift for a satisfactory description.
The capacitor model provides a meaningful measure to compare different surfaces with comparable
graphene-metal distances via the work function. In addition, the work function for graphene-covered 
surfaces can be estimated from the work function of the substrate at least for weakly coupled graphene.

The charges transferred between the substrate and the graphene layer originate from the Dirac cone
on the graphene side. These might be electron or holes depending on the doping of the graphene.
The shift of the work function due to the charge transfer can therefore be directly related 
to the energy of the Dirac point relative to the Fermi energy. 
Zero is obtained for a work function of 4.5~eV\@.
The values are taken from the literature \cite{Sutter2009,Bostwick2007,Marchenko2011,Niesner2013,Pletikosic2009} and are plotted in figure~\ref{fgr:workfun} as filled red circles. The agreement is perfect for SiC. but for the other surfaces larger deviations are observed. One has to keep in mind that for the noble-metal substrates the graphene
layer is p-doped and the Dirac point cannot be observed directly in photoemission spectroscopy. Its energy is extrapolated from the dispersion of the Dirac cone below the Fermi energy.
The experimental results are in good agreement with the calculations, which show the constant
difference between work function and energy of Dirac point as predicted by calculations \cite{Giovannetti2008,Khomyakov2009}. 
This has been found also for different modifications of graphene on SiC \cite{Gugel2013}.

\subsection{Binding energies}\label{binden}

The important parameter of the dielectric model is the charge density of the graphene layer
which can be related to the doping level and work function. The energies of the image-potential states
as a function of work function have been plotted as red solid lines in figure~\ref{fgr:EnWth}.
Large open circles (blue) show the available experimental values \cite{Niesner2012,Nobis2013,Gugel2013}. 
The symbol size represents typical experimental uncertainties.  
The data points for graphite were taken from\cite{takahashi12}
which observed the lowest three image-potential states. 
Other groups reported binding energies of $0.85\pm0.10$ eV for the $n=1$ image-potential 
state\cite{pagliara,yamamoto}.
The experimental values are slightly above the calculated lines, but the slope agrees fairly well. The agreement could be improved by using a smaller value
for the parameter $\alpha$ relating work function and Fermi-Thomas wave vector $W=\alpha k_\mathrm{FT}$. 
The data point for the $n=1$ state of SiC lies significantly too high.
This might be an effect of different screening properties of the dielectric substrate or residual
binding to the buffer layer. The accuracy and work function range of the experimental data for the noble-metal
surfaces is not sufficient to determine the slope independently. Therefore we only conclude that
the experimental data are in reasonable agreement with the predictions of the  dielectric model calculations.

Finally, we discuss the interpretation of experimental data measured on Gr/Ru \cite{hofer12}.
Three bound states have been measured with respect to the Fermi
energy (all energies in eV): $3.44$ (1'), $3.59$ (1) and $3.82$ (2).
The work function was measured as 4.24~eV\@. On the corrugated graphene
on Ru(0001) surface also lower areas exist which have a work function
of 4.00~eV\@. On all other surfaces only the $n=1$ and $n=3$ image-potential states
have been found. It is therefore worthwhile to check whether on Ru the additional state
might be the $n=2$ image-potential state. The data are plotted by green solid circles with error bars
in figure~\ref{fgr:EnWth} for the two different values of the work function.
For a work function of 4.00~eV the experimental binding energies agree reasonably well with the calculated lines. 
This assignment would also be compatible with the observed monotonous decrease of the lifetime with 
binding energy \cite{hofer12}. However, we cannot rule out 
the consistent interpretation based on different local work functions 
on the corrugated graphene on Ru as proposed by Armbrust {\it et al.}\ \cite{hofer12}.

\section{Conclusions}\label{concl}
Using standard models for the dielectric response and the reflection of electromagnetic
waves at a surface we have computed the static self-energy for an ultra-thin
slab mimicking a graphene layer. 
The self-induced potential goes continuously from the exchange and correlation energy
inside the material to the classical asymptotic image potential in the vacuum. 
For the purpose of obtaining image-potential states binding energies we find that FT makes
an excellent and convenient approximation to the accurate RPA.
Eigenvalues and eigenfunctions have been compared with Whittaker classical series
and recent experiments on Gr/Ir.
A free standing graphene ultra-thin layer produces a spatially symmetric self-energy
that induces an image potential series with even and odd states. 
The odd members of the series show a remarkable resemblance to 
the solution of Schr\"odinger equation for the
classical image potential (Whittaker wave functions).
On the other hand, even wave functions arise as new 
states that differ from Whittaker
in several key respects, e.~g. their non-zero 
density probability at the origin.
While the qualitative aspects of the image series supported by
a thin slab can be described in terms of quite general physical properties,
the detailed quantitative values depend crucially on the penetration of
wave functions into the substrate and the thin graphene film supported on it.
We have considered the limiting cases of full and none penetration.
For the case of films weakly interacting with a support and wave functions
penetrating well inside the system
some new states may consequently appear in between the classical ones, that
can be traced back to the even states in a free-standing
slab. 
In cases where the interaction is strong and the surface electronic structure
prevents the penetration of wave-functions a behavior more similar to a
standard metallic surface is expected.
We notice that 
the formalism used here can be easily generalized to include the effect of surface
plasmons via a frequency-dependent dielectric function \cite{sols87}.

The experimental results for graphene on
various substrates compare well to theoretical predictions.  The measured 
work function change due to graphene agree with the capacitor model of Giovanetti {\it et al.}\
\cite{Giovannetti2008,Khomyakov2009}. This opens the possibility to predict the work
function of graphene-covered surfaces from the substrate work function.
The work function difference from an isolated graphene sheet is related to the doping
of the graphene layer and determines in turn the available screening charge. The
resulting change in the energy of the image-potential states calculated with the theoretical model
is in agreement with the experimental data
from two-photon photoemission.

\ack
This work has been financed by the Governments of Spain
(MAT2011-26534, and FIS2010-19609-C01-01),
and the Basque Country (IT-756-13).
Computing resources provided by the CTI-CSIC are 
gratefully acknowledged.

\appendix

\section{Electrodynamics of a slab characterized by a width $2 d$ 
and $\epsilon(k)$}\label{Slab}

We are interested in the quasi-static
self-energy potential by an external charge near a slab 
characterized by a  dielectric function, $\epsilon(k)$. 
The dynamical problem, however, is solved exactly in the same
way by introducing a k and w-dependent response function $\epsilon(k,w)$. 
The problem is solved independently for two homogeneous systems
(pseudo-vacuum and pseudo-medium), with a set of fictitious
charges, $\sigma$, to reproduce the real fields in the regions for the vacuum
and the material respectively \cite{GMFF}.
Notice that the electromagnetic field propagator in vacuum, $\frac{4 \pi}{k^2}$,
corresponds to three spatial dimensions, 
$\vec k =(\vec \kappa,q)$.
Symmetric and antisymmetric solutions are analyzed separately and
combined to yield a solution for the most general case. 
The extra fictitious charges are finally removed from the solutions
by using matching conditions. 
Since a free standing graphene slab is symmetric
with respect to the middle plane, matching conditions on
only one surface need to be considered. 

\begin{figure}
\includegraphics[clip,width=0.8\columnwidth]{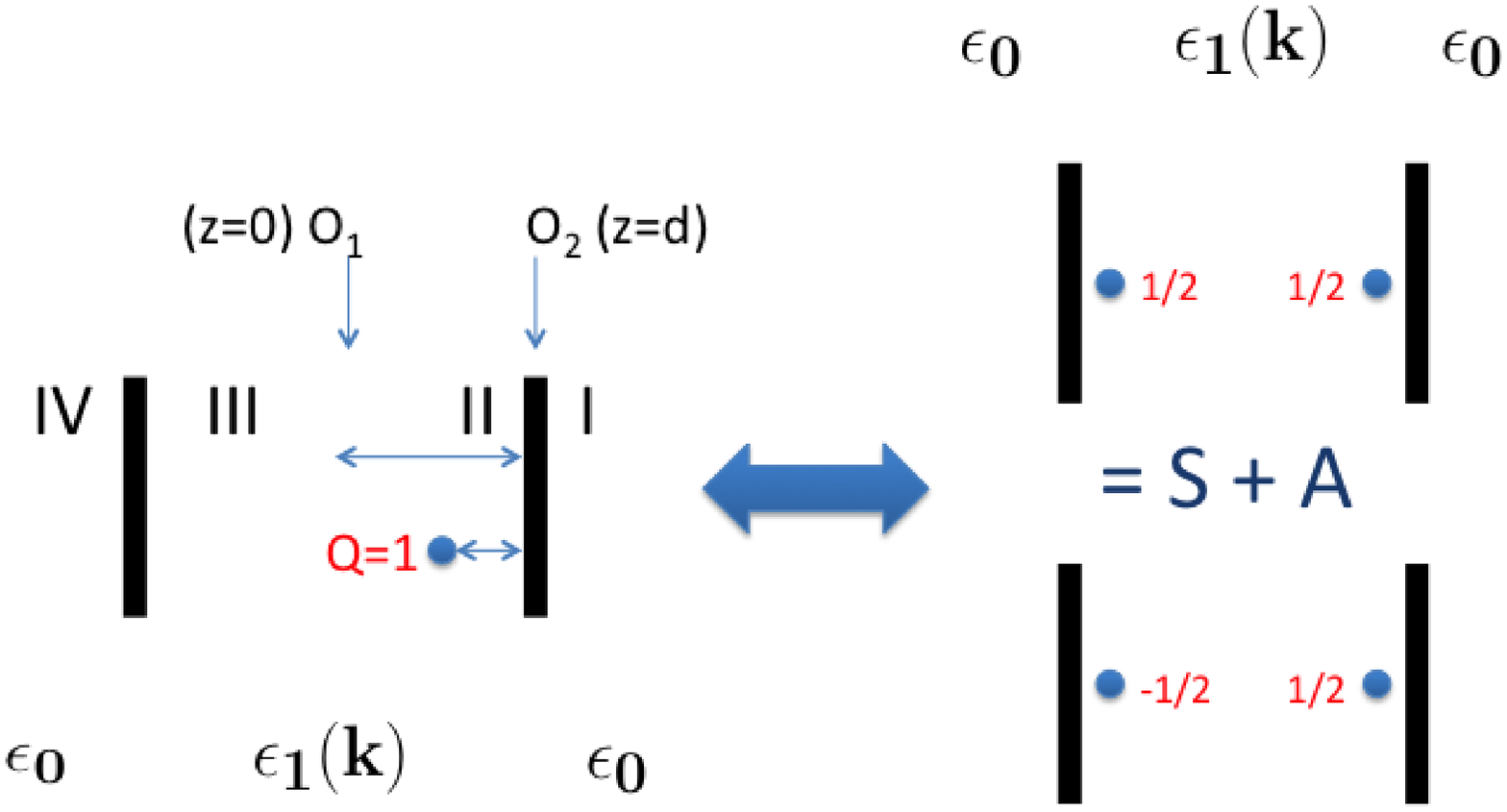}
\caption{
Q inside the ultra-thin slab ($0 \le z_1 \le d$, region II). 
The problem is decomposed as a superposition
of symmetric (S, upper part) 
and antisymmetric (A, lower part) configurations.
The slab polarization is characterized by $\epsilon(\vec k)$,
and its width, $2d$, 
and it is symmetric around O$_1$ ($z=0$).
Specular reflection conditions for the fields
are imposed at one of the two equivalent surfaces (O$_{2}$ $z=d$).
}
\label{fgr:QgrSA}
\end{figure}

The self-energy of the external probe charge, $Q$ is obtained
by computing the total potential in each region, and subtracting
the bare potential due to $Q$:
\begin{equation}
\Phi(z, z_1)
= \frac{1}{2}
\int \frac{d^2 k_{\parallel}}{(2 \pi)^2} \frac{dq}{2 \pi}
e^{i q z}
\lbrace
\phi^{S}(k) + \phi^{A}(k) - \frac{4 \pi Q}{k^2} e^{-i q z_1}
\rbrace
\label{eq:selfV}
\end{equation}
\noindent
where $\phi^{S,A}(k)$ must be separately obtained for Q
inside/outside the slab. The case where Q is inside the
material is schematically shown in figure~\ref{fgr:QgrSA}.

\subsection{Q inside the slab (II and III, $-d \le z \le d$)}

\begin{figure}
\includegraphics[clip,width=0.8\columnwidth]{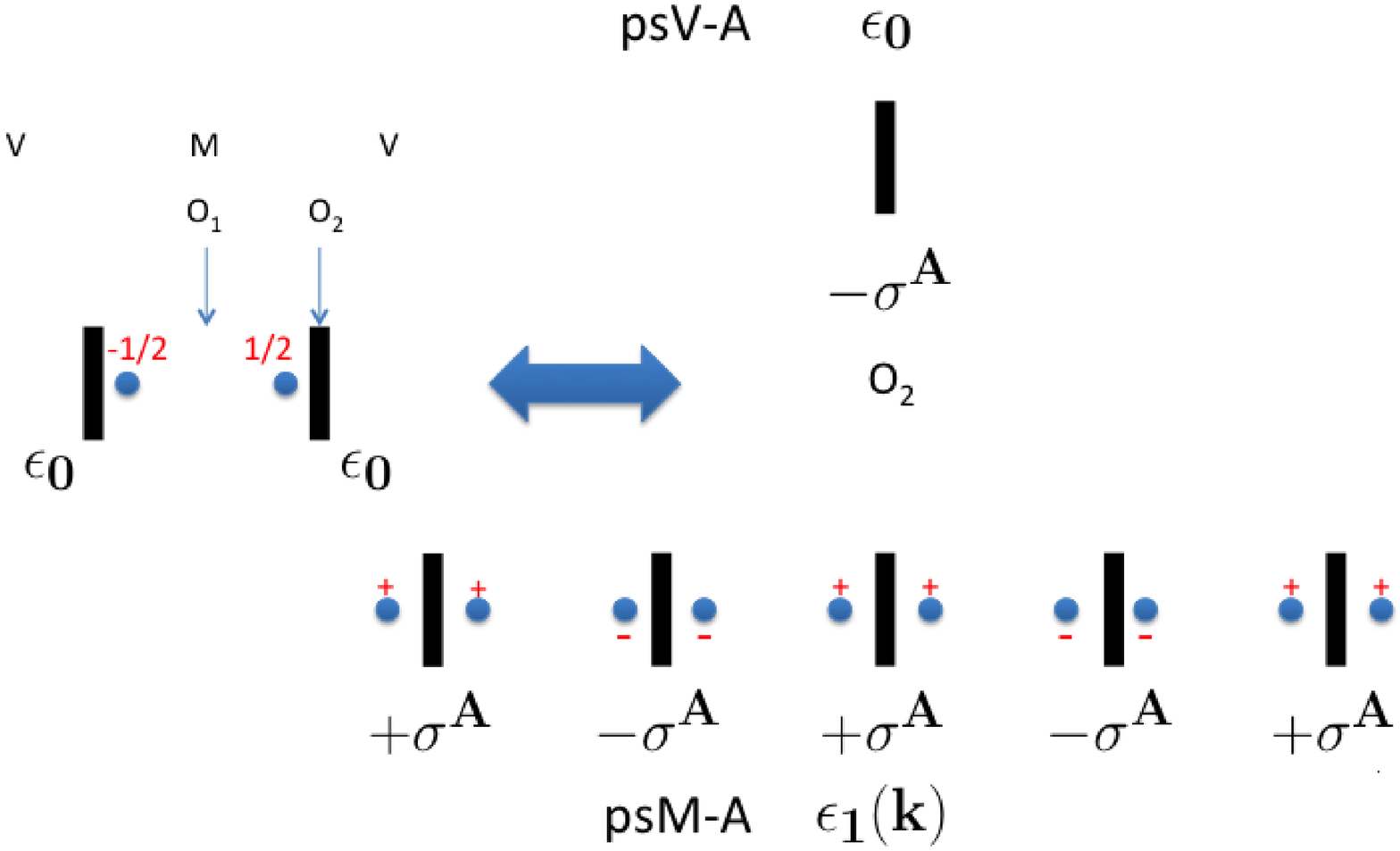}

\bigskip
\includegraphics[clip,width=0.8\columnwidth]{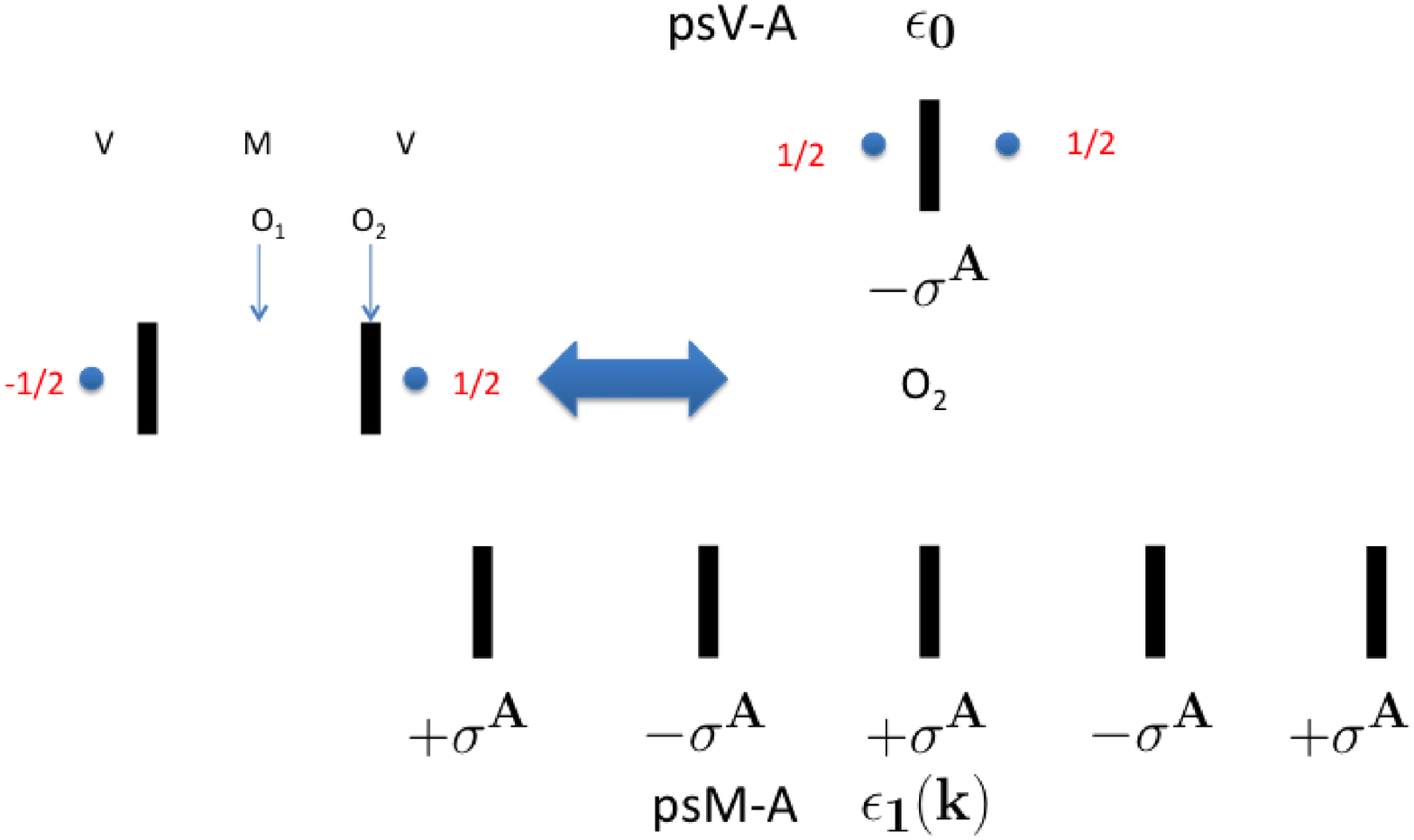}
\caption{
Schematics for the extended
pseudo-vacuum and pseudo-medium with charges distribution 
appropriate for the
antisymmetric case. 
Upper panel: $Q$ inside the slab. 
Lower panel: $Q$ in the vacuum region.
}
\label{fgr:QAnt}
\end{figure}

An schematic distribution for the pseudo-charges
when the external Q is inside the slab is given in
figure~\ref{fgr:QAnt} for the antisymmetric case.
The symmetric configuration follows easily by substituting 
the factors $(-1)^n$ in the sums by $(+1)^n$.
Continuity of the perpendicular component of the
displacement field results in $\sigma^{V}=-\sigma^{M}$ for
the pseudo-vacuum, $V$, and the pseudo-medium, $M$, for both
the symmetric and antisymmetric cases. 

Therefore, the total potential for the symmetric (antisymmetric) cases, are:

\begin{equation}
\phi^{V}(k) =-
\frac{4 \pi}{k^2}
\sigma_{\kappa}
\end{equation}

\[\hspace*{-9mm}
\phi^{M}(k) =
\frac{4 \pi}{k^2 \epsilon(k)}
\left(
\sigma_{\kappa}
\sum_{n=-\infty}^{\infty} (\pm)^n e^{i q 2 d n} 
+
Q
 \sum_{n=0}^{\infty} (\pm)^n 
 \left(
 e^{i q (z_1 + 2 d n)} \pm
 e^{i q (2 d + z_1 + 2 d n)}  
 \right)
\right)
\label{eq:II}
\]

The surface charges, $\sigma^{S}$ and $\sigma^{A}$, are separately obtained
from the condition:
$\Phi^{V}(z=d^{+})=\Phi^{M}(z=d^{-})$.

Convergence of the different series in these expressions
is guaranteed by the zeros of the dielectric function,
i.~e.\ the normal modes dressed by the interaction
in the medium \cite{GMFF}.
For the simple Thomas-Fermi dielectric function we have,
$k^2 \epsilon_\mathrm{FT}(k) = (q^2 + \kappa^2 + k_\mathrm{FT}^2) =
(q+i\chi)(q-i\chi)$, 
and the integration over the perpendicular moment, $q$,
can be performed to obtain 
analytic or semi-analytic expressions
that are given below. 

\subsection{Q inside the vacuum (I and IV, $\mid z \mid \ge d$)}

A similar procedure yields for the case of Q in the vacuum region:

\begin{equation}
\phi^{V}(k) =
\frac{4 \pi}{k^2}
\lbrace
-\sigma_{\kappa}
+
\frac{Q}{2} \left(
 e^{-i q z_1} -
 e^{i q z_1}  \right)
\rbrace
\end{equation}

\[
\phi^{M}(k) =
\frac{4 \pi}{k^2 \epsilon(k)}
\sigma_{\kappa}
\sum_{n=-\infty}^{\infty} (\pm 1)^n e^{i q 2 d n}
\]

Equation~(\ref{eq:selfV}) yields a numerical procedure that
allows to obtain the self-induced potential.
Using the FT approximation, quasi-explicit expressions 
that only depend on a single numerical integration,
have been obtained and given in (\ref{eq:FTvI}).
We remark this is an excellent approximation to the full
RPA result for the purpose of computing binding energies
of image-potential states.


\section{Useful analytical results in the
Fermi-Thomas approximation.}\label{Asymptotics}

The semi-infinite system ($d \to \infty$), along with the use
of Fermi-Thomas dielectric function, brings some simplifications
to the expression for the potential that can be exploited to compute
exact values at the surface, well inside the material, and
in the asymptotic vacuum region.
For convenience, in this appendix we move the origin to the surface ($z=d$).

\subsection{Induced potential in the vacuum region.}

Using the equations in \ref{Slab}
we obtain the induced potential in the vacuum  (now $z>0$):

\begin{equation}
\Phi(z>0) = 
\frac{1}{2} \int_{0}^{\infty} d \kappa e^{- \kappa 2 z}
\lbrack
1 - \frac{2}
{1+\frac{\kappa}{\pi} \int \frac{d q}{k^2 \epsilon(k)}}
\rbrack
\end{equation}
\noindent
that can be further simplified by the use of 
$k^2 \epsilon(k)=k^2 \epsilon_\mathrm{FT}(k)=\kappa^2 +q^2 + k^2_\mathrm{FT}$:
\begin{equation}
=
\frac{1}{2} \int_{0}^{\infty} d \kappa e^{- \kappa 2 z}
\frac{\kappa-\chi}{\kappa+\chi} =
\label{VvacFTscib}
\end{equation}
\[
-\frac{1}{12 z^3} 
\lbrace
6 ~ {\bf _{0} \tilde F_{1}}\left(;-1;-k_\mathrm{FT}^2
   z^2\right)-3 \pi  k_\mathrm{FT}^2 z^2 {\bf H}_2(2 k_\mathrm{FT}
   z)+k_\mathrm{FT}^2 z^2 (4 k_\mathrm{FT} z+3)+
\]
$$
     +\left(3-6
   k_\mathrm{FT}^2 z^2 \log (k_\mathrm{FT} z)\right) J_0(2 k_\mathrm{FT}
   z)+6 k_\mathrm{FT} z (\log (k_\mathrm{FT})+\log (z)+1) J_1(2
   k_\mathrm{FT} z)+3
   \rbrace
$$

\noindent
where ${\bf _{0} \tilde F_{1}}(;b;z)$ is the regularized confluent hypergeometric function,
${\bf H}_2(z)$ is the Struve function, and 
$J_{n}(z)$ are Bessel functions of the first kind.

\subsection{Asymptotic behaviour and position of the image plane.}

We can study the asymptotic behavior  in the long wave-length region ($\kappa \approx 0$)
by expanding the fraction in the integrand in Eq. \ref{VvacFTscib},
$\frac{\kappa-\chi}{\kappa+\chi} \approx -1 + \frac{2 \kappa}{k_\mathrm{FT}} + O(\kappa^2)$,
to obtain:

\begin{equation}
\Phi(z) \approx -\frac{1}{4z} + \frac{1}{4 k_\mathrm{FT} z^2} 
\end{equation}

On the other hand, we can expand the classical image potential around $z=\infty$

\begin{equation}
-\frac{1}{4(z-z_0)} \approx -\frac{1}{4z} -\frac{z_0}{4z^2} + O(\frac{1}{z^3})
\end{equation}

\noindent
which allows us to identify the position of the image plane
by direct comparison of both expressions, $z_0 = -\frac{1}{k_\mathrm{FT}}$. 

This asymptotic expression,
corrected by the position of the image plane obtained above, 
makes an excellent approximation
to the full potential either in the vacuum region outside the slab or
in the middle of a vacuum gap, the case discussed below.

\subsection{Particular values for $z=0$ and $z=-d$.}

For $z=0$ ($d\gg\frac{1}{k_\mathrm{FT}}$) we have:
 
\begin{equation}
\Phi(z=0) = 
\frac{1}{2} \int_{0}^{\infty}  
e^{- \kappa 2 z}
\frac{\kappa-\chi}{\kappa+\chi} d \kappa
= -\frac{k_\mathrm{FT}}{3}
\end{equation}

\noindent
For $z=-d$,

\begin{equation}\hspace*{-9mm}
\Phi(z=-d) = 
\frac{1}{2} \int \frac{d^{3} \vec k}{(2 \pi)^{3}} \frac{4 \pi}{k^2} 
\left( \frac{1}{\epsilon_\mathrm{FT}(k)}-1\right)
=
\frac{k_\mathrm{FT}^2}{2} \int_{0}^{\infty} d \kappa 
\frac{1}{\kappa \chi + \chi^2} = -\frac{k_\mathrm{FT}}{2}
\end{equation}

\subsection{Potential in a vacuum gap.}\label{VacuumGap}

It is handy to apply the same techniques to the inverse problem:
the potential in a vacuum gap between two semi-infinite media at $z< \pm d$.
The theoretical procedure proceeds along similar lines we have analyzed
in this paper, and we simply give the Fermi-Thomas result here:

\begin{equation}
\Phi(z>d)=
-\frac{k_\mathrm{FT}}{2}-\frac{e^{2 k_\mathrm{FT} (d-z)}}{4 (d-z)} -
\end{equation}

\[\hspace*{-2mm}
-
\int_{0}^{\infty} d \kappa \
\frac{  \kappa^2 e^{2 (d-z) \sqrt{\kappa^2+k_\mathrm{FT}^2}}  
\left(\kappa+\sqrt{\kappa^2+k_\mathrm{FT}^2} \coth[2 d \kappa]\right)}
{\sqrt{\kappa^2+k_\mathrm{FT}^2}
\left(\kappa+\sqrt{\kappa^2+k_\mathrm{FT}^2} \coth[d \kappa]\right) 
\left(\kappa+\sqrt{\kappa^2+k_\mathrm{FT}^2} \tanh[d \kappa]\right)}
\]

\noindent
and,

\begin{equation}
\Phi(-d \le z \le d)=
\end{equation}

\[
\int_{0}^{\infty} d \kappa \
e^{-2 \kappa z} \
\frac{
-k_\mathrm{FT}^2-e^{4 \kappa z} k_\mathrm{FT}^2+
e^{2 \kappa (-d+z)} \left(2 k_\mathrm{FT}^2+4 \kappa
\left(\kappa-\sqrt{\kappa^2+k_\mathrm{FT}^2}\right)\right)}
{4 \left(2 \kappa \sqrt{\kappa^2+k_\mathrm{FT}^2} 
\cosh[2d \kappa]+\left(2 \kappa^2+k_\mathrm{FT}^2\right) 
\sinh[2 d \kappa]\right)}
\]

\begin{figure}
\includegraphics[clip,width=0.9\columnwidth]{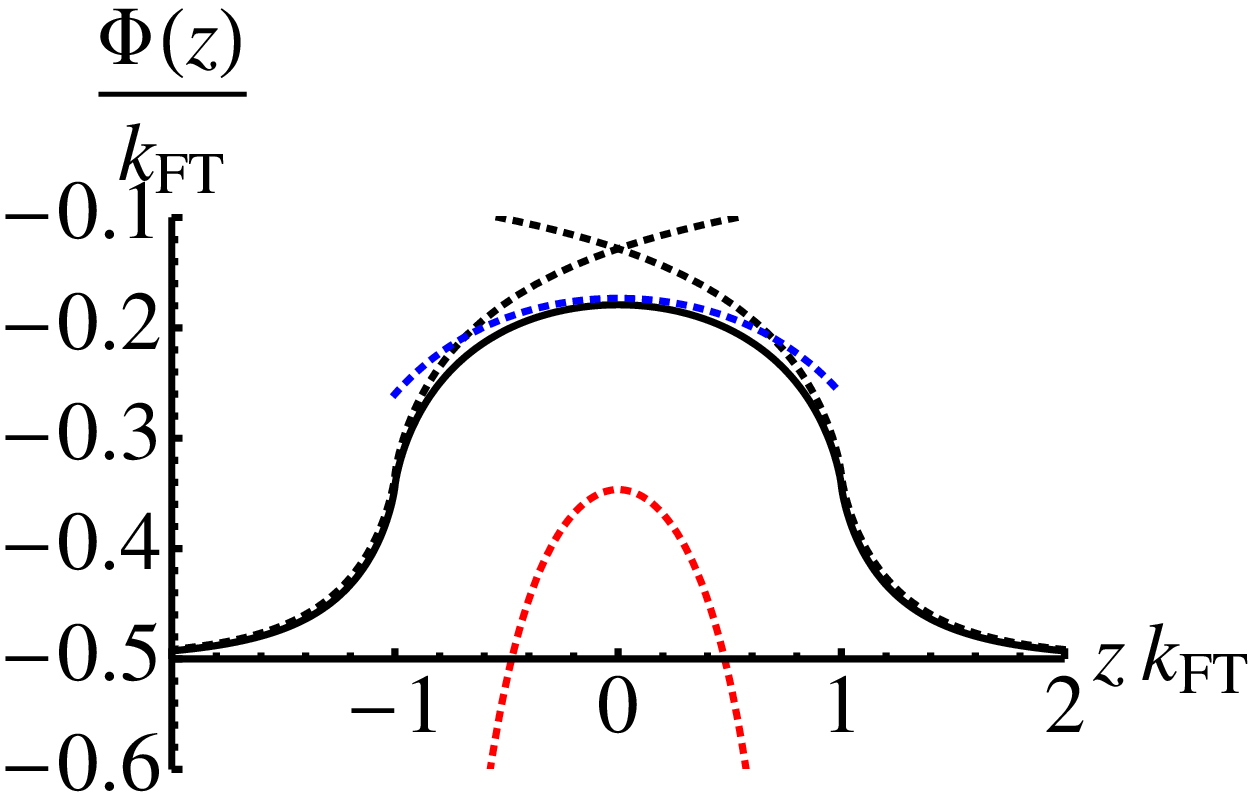}
\caption{
Self-induced potential, $\Phi(z)$, of a unit test charge at different positions
in a vacuum gap of width 
$2$ a.u., and inside the surrounding metals ($k_\mathrm{FT}=1$ a.u.).
Continuous line (black): non-local potential for the gap (Fermi-Thomas). 
Dashed line (black): non-local potential for each surface in the gap
considered as independent systems (Fermi-Thomas).
Dashed line (red): classical result taking into account all the 
classical images and counter-images \cite{sols87}.
Dashed line (blue): semi-classical result adding the contributions
of all the classical images, but corrected by an image plane
at $z_0=\pm \frac{1}{k_\mathrm{FT}}$ (valid in the vacuum region only).
}
\label{fgr:GAP31}
\end{figure}

This potential has been plotted in
figure~\ref{fgr:GAP31},
where it is compared
with: (i) the classical value \cite{sols87,hofer12b}, 
  
\begin{equation}
\Phi(|z|\le d)=
-\frac{2 \gamma-
\Psi^{(0)}\left(\frac{1}{2} +\frac{z}{2 d}\right)-
\Psi^{(0)}\left(\frac{1}{2} -\frac{z}{2 d}\right)}
{8 d}
\end{equation}
\noindent
($\Psi^{0}(z)$ is the digamma function, and $\gamma=-0.577$ is the
Euler constant),
(ii) the classical value
corrected by an image plane determined by an expansion
to the full non-local potential calculated in the Fermi-Thomas
approximation, and (iii) the independent non-local potential for
each of the two surfaces delimiting the vacuum gap.

\clearpage
\newpage


\end{document}